# Platinum/Yttrium Iron Garnet Inverted Structures for Spin Current Transport

Mohammed Aldosary[1], Junxue Li[1], Chi Tang[1], Yadong Xu[1], Jian-Guo Zheng[2], Krassimir N. Bozhilov[3], and Jing Shi[1]

[1]Department of Physics and Astronomy and SHINES Energy Frontier Research Center, University of California, Riverside, CA 92521, USA.

[2]Irvine Materials Research Institute, University of California, Irvine, CA 92697, USA.

[3]Central Facility for Advanced Microscopy and Microanalysis, University of California, Riverside, CA 92521, USA.

30-80 nm thick yttrium iron garnet (YIG) films are grown by pulsed laser deposition on a 5 nm thick sputtered Pt atop gadolinium gallium garnet substrate (GGG) (110). Upon post-growth rapid thermal annealing, single crystal YIG(110) emerges as if it were epitaxially grown on GGG(110) despite the presence of the intermediate Pt film. The YIG surface shows atomic steps with the root-mean-square roughness of 0.12 nm on flat terraces. Both Pt/YIG and GGG/Pt interfaces are atomically sharp. The resulting YIG(110) films show clear in-plane uniaxial magnetic anisotropy with a well-defined easy axis along <001> and a peak-to-peak ferromagnetic resonance linewidth of 7.5 Oe at 9.32 GHz, similar to YIG epitaxilly grown on GGG. Both spin Hall magnetoresistance and longitudinal spin Seebeck effects in the inverted bilayers indicate excellent Pt/YIG interface quality.

Magnetic garnets are important materials that offer unique functionalities in a range of bulk and thin film device applications requiring magnetic insulators.[1,2] Among all magnetic insulators, yttrium iron garnet ($Y_3Fe_5O_{12}$ or YIG) has been most extensively used in various high-frequency devices such as microwave filters, oscillators, and Faraday rotators[3] due to its attractive attributes including ultra-low intrinsic Gilbert damping constant ($\alpha$ as low as $3 \times 10^{-5}$)[4] which is two orders of magnitude smaller than that of ferromagnetic metals, high Curie temperature ($T_C$ = 550 K), soft magnetization behavior, large band gap (~ 2.85 eV),[5] and relatively easy synthesis in single crystal form. These conventional applications demand bulk YIG crystals or micron-thick films grown by liquid phase epitaxy.[6] For more recent spintronic studies such as the spin Seebeck effect (SSE)[7] and spin pumping,[8] submicron- or nanometer-thick films are typically grown by pulsed laser deposition (PLD) or sputtering. It has been shown that high-quality YIG films can be epitaxially grown directly on GGG substrates due to the same crystalline structure and a very small lattice mismatch of 0.057%.[9-11] To form bilayers, a thin polycrystalline metal layer is typically deposited on top of YIG by sputtering, which results in reasonably good interfaces for spin current transport.[7,8,12] For some studies such as the magnon-mediated current drag,[13,14] sandwiches of metal/YIG/metal are required, in which YIG needs to be both magnetic and electrically insulating. However, high-quality bilayers of the reverse order, i.e. YIG on metal, are very difficult to be fabricated. A main challenge is that the YIG growth requires high temperatures and an oxygen environment [15] which can cause significant inter-diffusion, oxidation of the metal layer, etc. and consequently lead to poor structural and electrical properties in both metal and YIG layers.

This letter reports controlled growth of high-quality single crystal YIG thin films ranging from 30 to 80 nm in thickness on a 5 nm thick Pt layer atop $Gd_3Ga_5O_{12}$ or GGG (110) substrate. Combined with low-temperature growth which suppresses the inter-diffusion, subsequent rapid thermal annealing (RTA) and optimization of other growth parameters result in well-defined magnetism, atomically sharp Pt/YIG interface, and atomically flat YIG surface. In addition, despite the intermediate Pt layer that has a drastically different crystal structure from the garnets, the top YIG layer shows desired structural and magnetic properties as if it were epitaxially grown on GGG (110).

5 × 5 mm$^2$ of commercial GGG (110) single crystal substrates are first cleaned in ultrasonic baths of acetone, isopropyl alcohol, then deionized water, and dried by pure nitrogen gun. Subsequently, the substrates are annealed in a furnace at 900 °C in $O_2$ for eight hours which produces atomically flat surface. Atomic force microscopy (AFM) is performed to track the surface morphology of the annealed substrates. Figure 1(a) shows the 2x2 μm$^2$ AFM scan of an annealed GGG (110) substrate. Flat atomic terraces are clearly present and separated with a step height of 4.4 ± 0.2 Å which is equal to ¼ of the face diagonal of the GGG unit cell or the (220) interplanar distances of 4.4 Å of GGG. The 4.4 Å distance is the separation between the $GaO_6$ octahedral layers parallel to (110) that might be defining the observed atomic step ledges. The root-mean-square (RMS) roughness on the terraces is ~0.74 Å. Then, the substrate is transferred into a sputtering chamber with a base pressure of $5 \times 10^{-8}$ Torr for Pt deposition. DC magnetron sputtering is used with the Ar pressure of 5 mTorr and power of 37.5 W. The sputtering deposition rate is 0.76 Å/s and sample holder rotation speed is 10 RPM. After the 5 nm thick Pt deposition, the surface of the Pt film is found to maintain the atomic terraces of the GGG (110) substrate, except that the RMS roughness on the Pt terraces is increased to 1.05 Å as shown in figure 1(b). It is rather surprising that the 5 nm thick Pt layer does not smear out the terraces separated by atomic distances given that the sputtering deposition is not particularly directional. Strikingly, terraces are still present even in 20 nm thick Pt (not shown). The substrates are then put in a PLD chamber which has a base pressure of $4 \times 10^{-7}$ Torr, and are slowly heated to 450 °C in high-purity oxygen with the pressure of 1.5 mTorr with 12 wt% of ozone. The krypton fluoride (KrF) coherent excimer laser ($\lambda$ = 248 nm, 25 ns/pulse) used for deposition has a pulse energy of 165 mJ/pulse, and repetition rate of 1 Hz. The deposition rate of $\approx$ 1.16 Å/min is achieved with a target to substrate distance of 6 cm. After deposition, the YIG films are *ex situ* annealed at 850 °C for 200 seconds using rapid thermal annealing (RTA) under a steady flow of pure oxygen. After RTA, the surface morphology is examined by AFM again. Figure 1(c) shows the atomically terraced surface of a 40 nm thick YIG film with RMS of 1.24 Å on the terrace. In this study, the thickness of YIG ranges from 30 − 80 nm and all samples exhibit clear atomic terraces. Even though YIG is annealed at such a high temperature, with the short annealing time, the flat and smooth YIG surface is maintained.

To track the structural properties of YIG, we use RHEED to characterize the YIG surface at every step of the process. Figure 1(d) shows the RHEED pattern of the as-grown YIG surface. It clearly indicates the absence of any crystalline order. After the *ex situ* RTA, the sample is introduced back to the PLD chamber for RHEED measurements again. A streaky and sharp RHEED pattern is recovered as displayed in Figure 1(e) which suggests a highly crystalline order. This result is particularly interesting since it shows the characteristic RHEED pattern of YIG grown on GGG.[10]

To further confirm its crystalline structure, x-ray diffraction (XRD) using the Cu K$\alpha_1$ line has been carried out over a wide angle range (2$\theta$ from 10 to 90°) on the GGG/Pt/YIG sample discussed in Figure 2(a). Because of the close match in lattice constants between YIG and GGG substrate, weak YIG peaks are completely overlapped with strong peaks of GGG so that they are indistinguishable. Three main Bragg peaks of YIG and GGG are observed: 220, 440, and 660, which suggests the (110) growth orientation of both YIG and GGG. No individual weak YIG peaks can be found. It is striking that the YIG film adopts the crystallographic orientation of GGG despite the intermediate Pt layer. By comparing with the spectra of YIG grown directly on GGG, we can identify a new peak (2θ ≈ 40.15°) which is better seen in the zoom-in view in the inset of Figure 2(a). We determine this as the 111 peak of the 5 nm thick Pt film that suggests the (111) texture of the Pt layer. It is not clear whether the (111) texture in the intermediate Pt layer is required for YIG to develop the same crystallographic orientation as that of the GGG substrate.

The locking of the (110) orientation in both YIG and GGG is further investigated by the high-resolution transmission electron microscopy (HRTEM) in real space. Figure 2(b) first reveals sharp and clean interfaces of Pt/YIG and GGG/Pt. No amorphous phase or inclusions are visible at these two interfaces. Furthermore, the (110) atomic planes of YIG and GGG are parallel to each other and show very closely matched inter-planar spacing. Despite the Pt layer in between, the crystallographic orientation of YIG is not interrupted as if it were epitaxially grown on GGG directly. In the selected area electron diffraction pattern shown in figure 2(c), taken along the <112> zone axis in garnet from an area that includes all three phases, YIG and GGG diffraction spots overlap with each other, consistent with the XRD results. There is minor splitting of the 110 type reflections from the two garnet phases due to a slight rotation of the two garnet lattices of less than 0.5°. Surprisingly, the diffraction spots from the 5 nm Pt layer show a single crystal pattern with minor streaking parallel to 111 in Pt. The diffuse character of the Pt reflections suggests that Pt is essentially a single crystal consisting of small (few nanometers) structural domains with minor misalignments. The contrast variation in different regions of Pt shown in figure 2(b) is consistent with such small structural domain misalignments in Pt crystal grain orientations. Furthermore, the 111 reciprocal vector of Pt and the 110 reciprocal vector of YIG/GGG are both perpendicular to the interfaces, indicating that the (111) Pt layers are parallel to the (110) layers of both GGG and YIG. Figure 2(d) is a HRTEM image with high magnification of the three layers. It further reveals atomically sharp interfaces, interlocked (110) crystallographic orientations between GGG and YIG, and single crystal (111)-oriented Pt.

To investigate the magnetic properties of the GGG/Pt/YIG inverted heterostructure, vibrating sample magnetometry (VSM) measurements are carried out at room temperature. As-grown YIG films do not show any well-defined crystalline structure as indicated by the RHEED pattern. In the meantime, the VSM measurements do not show any detectable magnetization signal. Upon RTA, single crystal YIG becomes magnetic as shown by the hysteresis loops in Figure 3(a) for magnetic fields parallel and perpendicular to the sample plane. GGG's paramagnetic contribution has been removed by subtracting the linear background from the raw data. The easy axis of all YIG films with different thicknesses lies in the film plane due to the dominant shape anisotropy. The coercivity falls in the range of 15 - 30 Oe for different thicknesses, which is larger than the typical value (0.2 to 5 Oe)[9-11] for YIG films grown on lattice-matched GGG. The inset of Figure 3(a) shows a coercive field of 29 Oe for a 40 nm thick YIG film. The saturation magnetic field in the perpendicular direction is ~1800 Oe which corresponds well to $4\pi M_s$ for bulk YIG crystals (1780 Oe). Magnetic hysteresis loops are measured along different directions in the film plane. Figures 3(b & c) show the polar angular dependence of both the coercively field ($H_c$) and squareness ($M_r/M_s$) where $M_r$ is the remanence and $M_s$ is the saturation magnetizations, respectively. In the film plane, there is clear uniaxial magnetic anisotropy, with the in-plane easy and hard axes situated along <001> at φ = 145° and <110> at φ = 55°, respectively. This two-fold symmetry indicates that the magneto-crystalline anisotropy is the main source of the anisotropy since it coincides with the lattice symmetry of (110) surface of the YIG films, which is also consistent with the magnetic anisotropy property of YIG epitaxially grown on GGG (110).[10]

Ferromagnetic resonance (FMR) measurements of YIG films are carried out using Bruker EMX EPR (Electron Paramagnetic Resonance) spectrometer with an X-band microwave cavity operated at the frequency of f = 9.32 GHz. A static magnetic field is applied parallel to the film plane. Figure 3(d) shows a single FMR peak profile in the absorption derivative. From the Lorentzian fit, the peak-peak linewidth ($\Delta H_{pp}$) and resonance frequency ($H_{res}$) are 7.5 Oe and 2392 Oe, respectively. In literature, both the linewidth and the saturation magnetization vary over some range depending on the quality of YIG films. These values are comparable with the reported values for epitaxial YIG films grown directly on GGG.[9-11] The FMR linewidth here seems to be larger than what is reported in the best YIG films grown on GGG. Considering the excellent film quality, it is reasonable to assume that the same YIG would have similar FMR linewidth, e.g. 3 Oe. In the presence of Pt, increased damping in Pt/YIG occurs due to spin pumping.[16,17] This additional damping can explain the observed FMR linewidth (7.5 Oe) if a reasonable spin mixing conductance value of $g_{eff}^{\uparrow\downarrow} \approx 5 \times 10^{18} \, m^{-2}$ is assumed.

The Pt layer underneath YIG allows for pure spin current generation and detection just as when it is placed on top. It is known that the interface quality is critical to the efficiency of spin current transmission.[18,19] To characterize this property, we perform spin Hall magnetoresistance (SMR) and SSE measurements in GGG/Pt/YIG inverted heterostructures.

SMR is a transport phenomenon in bilayers of heavy metal/magnetic insulator.[12,20,21] A charge current flowing in the normal metal with strong spin-orbit coupling generates a spin current orthogonal to the charge current via the spin Hall effect. The reflection and absorption of this spin current at the interface of the normal metal/magnetic insulator depends on the orientation of the magnetization (**M**) of the magnetic insulator. Due to the spin transfer torque mechanism, when **M** is collinear with the spin polarization **σ**, reflection of the spin current is maximum. In contrast, when **M** is perpendicular to **σ**, absorption is maximum; therefore, the resistance of the normal metal is larger than that for **M** ∥ **σ** since the absorption behaves as an additional dissipation channel. Metal/magnetic insulator interface quality affects the SMR magnitude. As illustrated in figure 4(a), we carry out angle-dependent magnetoresistance (MR) measurements by rotating a constant magnetic field in the xy- (H=2000 Oe), xz- (H=1 T), or yz-plane (H=1 T), while the current flows along the x-axis. The angular dependence of the MR ratio, $\frac{\Delta\rho}{\rho}$ (%) = $\frac{\rho(angle)-\rho(angle=\frac{\pi}{2})}{\rho(angle=\frac{\pi}{2})} \times 100$, for Pt film at room temperature is summarized in figure 4(b). According to the SMR theory,[21] the longitudinal resistivity reads

$$\rho = \rho_0 + \rho_1 m_y^2 \qquad (1),$$

where $\rho_0$ and $\rho_1$ are magnetization-independent constants, and $m_y$ is the y-component of the magnetization unit vector. The red solid curves in figure 4(b) can be well described by equation (1). Here, the magnitude of SMR in xy- and yz- scans is on the same order as that in normal YIG/Pt bilayer systems. Therefore, we demonstrate that the SMR mechanism dominates in our devices, which indicates excellent interface quality for spin current transport.

SSE, on the other hand, is related to the transmission of thermally excited spin currents through the heavy metal/YIG interface.[22-24] As illustrated in figure 4(c), we first deposit a 300 nm thick $Al_2O_3$ layer atop GGG(110)/Pt(5 nm)/YIG(40 nm), and a top heater layer consisting of 5 nm Cr and 50 nm Au. When an electrical current (50 mA) flows in the Cr/Au layer, a temperature gradient is established along the z-direction by joule heating, which generates a spin current in YIG. As the spin current enters the Pt layer, it is converted into a charge current or voltage due to the inverse spin Hall effect. A magnetic field is applied in the y-direction while the voltage is detected along the x-direction. In Figure 4(d), we plot the field dependence of the normalized SSE signal at 300 K, which is consistent with the SSE magnitude reported in YIG/Pt bilayers[24]. Therefore, we have confirmed the excellent interface quality for transmitting thermally excited spin currents.

In summary, single crystal YIG thin films have been grown on Pt film which is sputtered on GGG (110) substrate. RHEED and AFM show excellent YIG surface quality and morphology. XRD and HRTEM further reveal an intriguing crystal orientation locking between YIG and GGG as if no Pt were present. These YIG films exhibit similar excellent magnetic properties to those of the YIG films grown epitaxially on GGG (110). Both SMR and SSE results confirm that the superb structural and magnetic properties lead to excellent spin current transport properties.

**ACKNOWLEDGMENTS**

We would like to thank Prof. J. Garay and N. Amos for the technical assistance and fruitful discussions. Bilayer growth control, growth characterization, device fabrication and electrical transport measurements at UCR were supported as part of the SHINES, an Energy Frontier Research Center funded by the U.S. Department of Energy, Office of Science, Basic Energy Sciences under Award No. SC0012670. Part of the transmission electron microscopy was performed on a 300 kV FEI Titan Themis at the Central Facility for Advanced Microscopy and Microanalysis at UC Riverside, supported by UCR campus funding. The TEM specimen preparation was performed at the Irvine Materials Research Institute (IMRI) at UC Irvine, using instrumentation funded in part by the National Science Foundation Center for Chemistry at the Space-Time Limit under Grant No. CHE-0802913.

# REFERENCES

[1] G. Winkler, Magnetic Garnets (Vieweg, Braunschweig, Wiesbaden, 1981).

[2] S. Geller and M. A. Gilleo, Acta Crystallogr. **10**, 239 (1957).

[3] A. V. Chumak, A. A. Serga, and B. Hillebrands, Nat. Commun. **5**, 4700 (2014).

[4] M. Sparks, Ferromagnetic-Relaxation Theory (Mc Graw- Hill, New York, 1964).

[5] X. Jia, K. Liu, K. Xia, and G. E. Bauer, Europhys. Lett. **96**, 17005 (2011).

[6] R. C. Linares, R. B. Graw, and J. B. Schroeder, J. Appl. Phys. **36**, 2884 (1965).

[7] D. Qu, S. Y. Huang, J. Hu, R. Wu, and C. L. Chien, Phys. Rev. Lett. **110**, 067206 (2013).

[8] B. Heinrich, C. Burrowes, E. Montoya, B. Kardasz, E. Girt, Y. Y. Song, Y. Sun, and M. Wu, Phys. Rev. Lett. **107**, 066604 (2010).

[9] M. C. Onbasli, A. Kehlberger, D. H. Kim, G. Jakob, M. Klaui, A. V. Chumak, B. Hillebrands, and C. A. Ross, APL Mater. **2**, 106102 (2014).

[10] C. Tang, M. Aldosary, Z. Jiang, H. Chang, B. Madon, K. Chan, M. Wu, J. E. Garay, and J. Shi, Appl. Phys. Lett. **108**, 102403 (2016).

[11] H. Chang, P. Li, W. Zhang, T. Liu, A. Hoffmann, L. Deng, and M. Wu, IEEE Magn. Lett. **5**, 6700104 (2014).

[12] T. Lin, C. Tang, H. M. Alyahayaei, and J. Shi, Phys. Rev. Lett. **113**, 037203 (2014).

[13] S. S.-L. Zhang and S. Zhang, Phys. Rev. Lett. **109**, 096603 (2012).

[14] J. Li, Y. Xu, M. Aldosary, C. Tang, Z. Lin, S. Zhang, R. Lake, and J. Shi, Nat. Commun. **7**, 10858 (2016).

[15] Y. Krockenberger, H. Matsui, T. Hasegawa, M. Kawasaki, and Y. Tokura, Appl. Phys. Lett. **93**, 092505 (2008).

[16] C. Burrowes, B. Heinrich, B. Kardasz, E. A. Montoya, E. Girt, Y. Sun, Y.-Y. Song, and M. Wu, Appl. Phys. Lett. 100, 092403 (2012).

[17] J. Lustikova, Y. Shiomi, Z. Qiu, T. Kikkawa, R. Iguchi, K. Uchida, and E. Saitoh, J. Appl. Phys. 116, 153902 (2014).

[18] M. Weiler, M. Althammer, M. Schreier, J. Lotze, M. Pernpeintner, S. Meyer, H. Huebl, R. Gross, A. Kamra, J. Xiao, Y.-T. Chen, H. J. Jiao, G. E. W. Bauer, and S. T. B. Goennenwein, Phys. Rev. Lett. **111**, 176601 (2013).

[19] Y. M. Lu, J. W. Cai, S. Y. Huang, D. Qu, B. F. Miao, and C. L. Chien, Phys. Rev. B **87**, 220409 (2013).

[20] H. Nakayama, M. Althammer, Y.-T. Chen, K. Uchida, Y. Kajiwara, D. Kikuchi, T. Ohtani, S. Geprägs, M. Opel, S. Takahashi, R. Gross, G. E. W. Bauer, S. T. B.Goennenwein, and E. Saitoh, Phys. Rev. Lett. **110**, 206601 (2013).

[21] Y.-T. Chen, S. Takahashi, H. Nakayama, M. Althammer, S. T. B. Goennenwein, E. Saitoh, and G. E. W. Bauer, Phys. Rev. B **87**, 144411 (2013).

[22] M. Schreier, A. Kamra, M. Weiler, J. Xiao, G. E. W. Bauer, R. Gross, and S. T. B. Goennenwein, Phys. Rev. B **88**, 094410 (2013).

[23] S. M. Rezende, R. L. Rodr´ıguez-Suarez, J. C. Lopez Ortiz, and A. Azevedo, Phys. Rev. B **89**, 134406 (2014).

[24] D. Meier, D. Reinhardt, M. van Straaten, C. Klewe, M. Althammer, M. Schreier, S. T. B. Goennenwein, A. Gupta, M. Schmid, C. H. Back, J.-M. Schmalhorst, T. Kuschel, and G. Reiss; Nat. Commun. **6**, 8211 (2015).

**Figure 1**

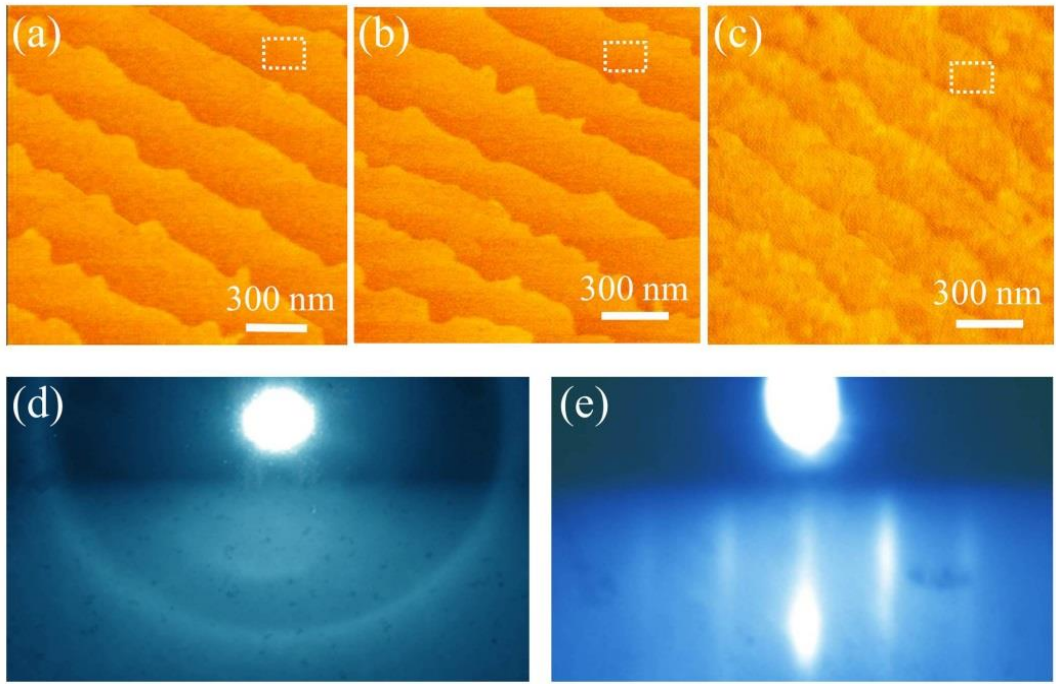

**FIG. 1. Surface characterization of YIG thin film grown on GGG(110)/Pt (5 nm)**. (a)–(c) 2 μm × 2 μm AFM scans of GGG(110) substrate, GGG(110)/Pt(5 nm), and GGG/Pt(5 nm)/YIG(40 nm), respectively. RHEED patterns of as-grown (d) and annealed (f) GGG(110)/Pt (5 nm)/YIG(40 nm).

**Figure 2**

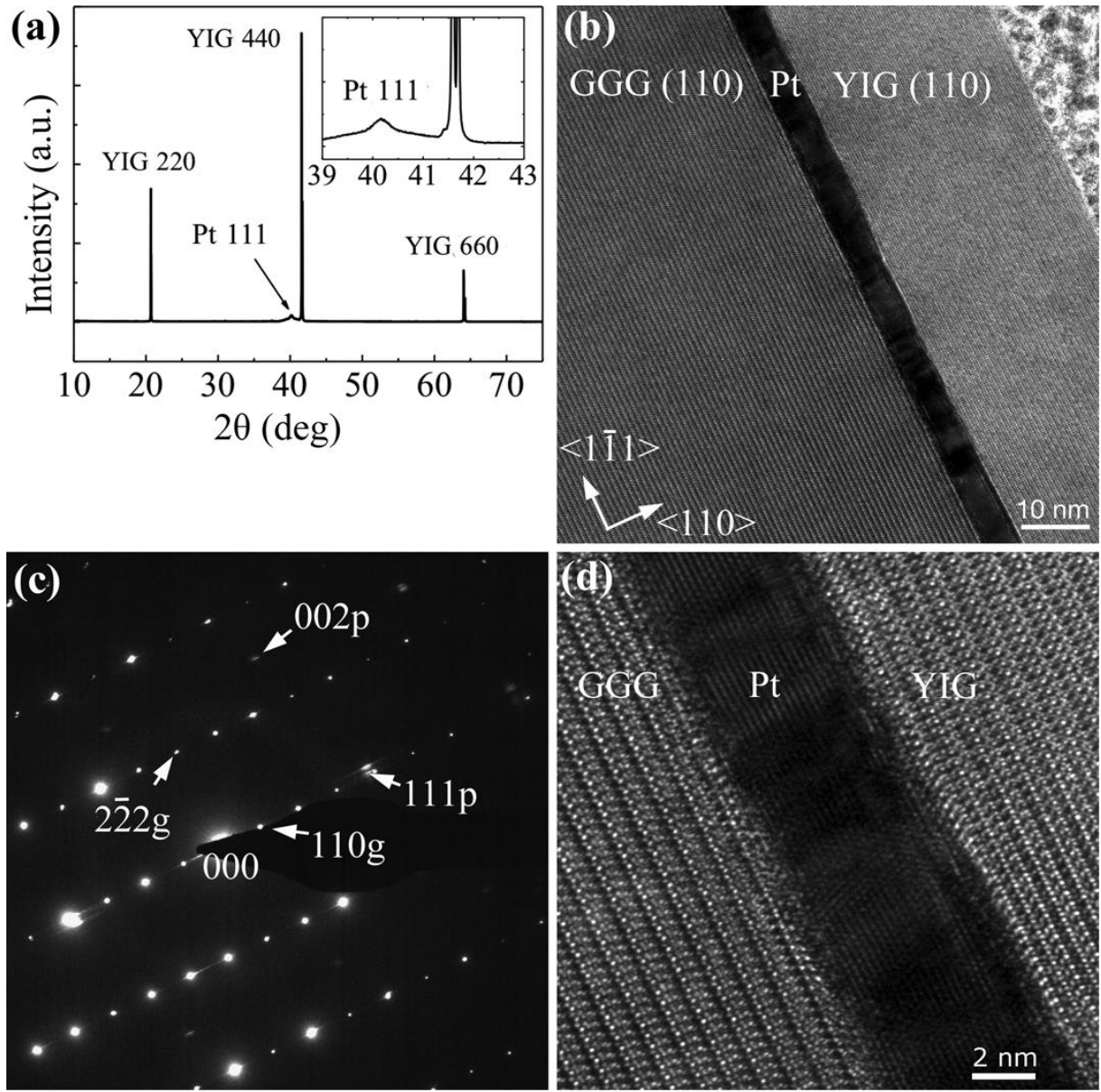

**FIG. 2. Structure characterization of GGG/Pt/YIG heterostructure.** (a) XRD of YIG film (40 nm) grown on GGG(110)/Pt (5 nm). Inset: zoom-in plot of Pt 111 peak (2θ = 40.15°). (b) TEM image of GGG (110)/Pt (5 nm)/YIG (110) (40 nm) heterostructure. The $<1\bar{1}1>$ and <110> directions in GGG are shown for reference. (c) Selected area electron

diffraction pattern along [$\bar{1}12$] zone axis in GGG obtained from an area containing all three layers showing diffraction spots of YIG, GGG and Pt. The garnet reflections are labeled with subscript "g" and Pt ones with "p". (d) HRTEM lattice image along the [$\bar{1}12$] zone axis in garnet shows that (110) planes in both YIG and GGG are parallel to the interface with the Pt film, and the latter is composed of nanometer size crystalline domains oriented with their (111) lattice planes parallel to the interface as well. Slight bending and disruption of the (111) lattice fringes between adjacent Pt domains are visualized.

**Figure 3**

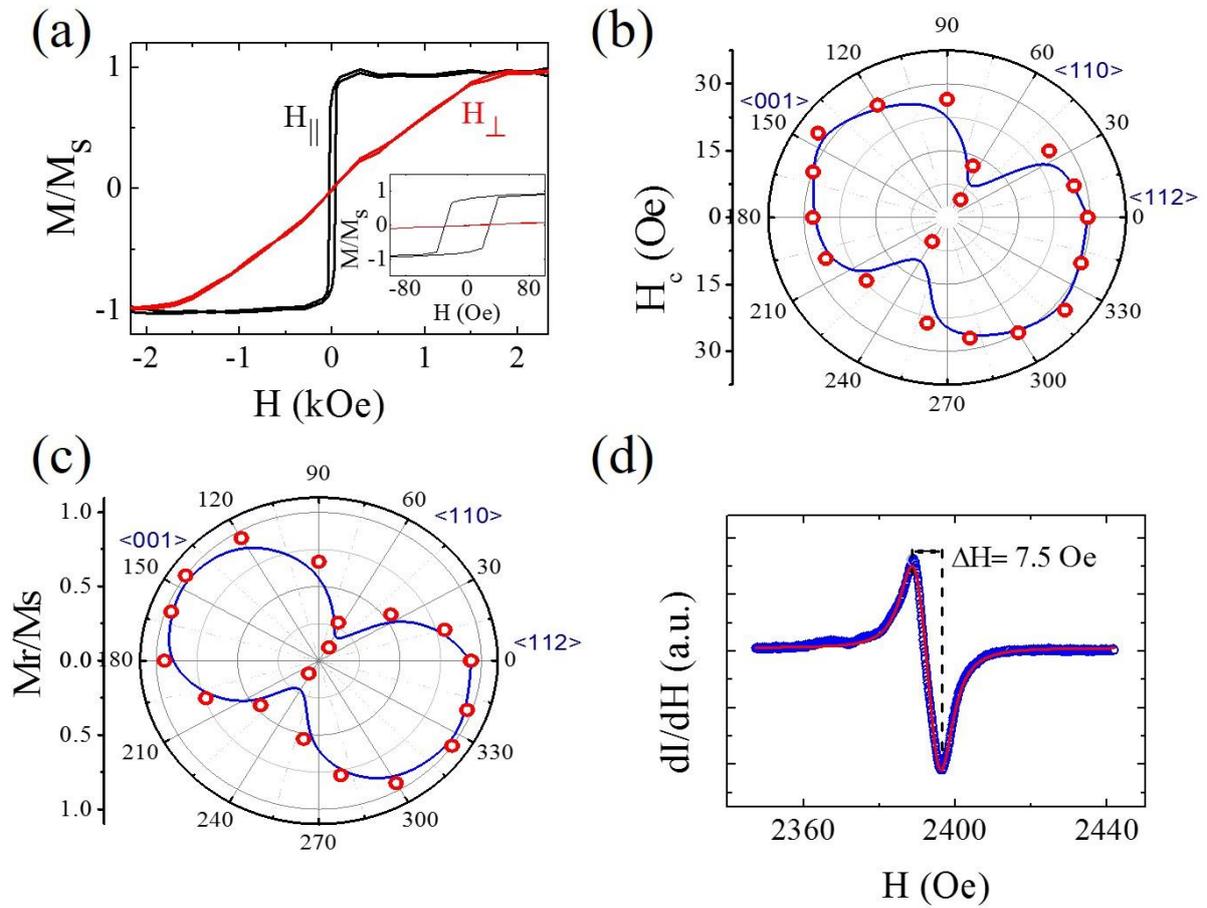

**FIG. 3. Magnetic properties of GGG(110)/Pt(5 nm)/YIG (40 nm)** (a) Room temperature normalized magnetic hysteresis loops of YIG (40 nm)/ Pt (5nm)/GGG (110) with magnetic field applied in-plane and out-of-plane. Inset: in-plane hysteresis loop at low fields. Polar plots of coercive field $H_c$ (b) and squareness $M_r/M_s$ (c) as the magnetic field **H** is set in different orientations in the (110) plane (**H** // <112> at 0°). (d) FMR absorption derivative spectrum of YIG/Pt/GGG at an excitation frequency of 9.32 GHz. Lorentzian fit (red line) shows a single peak with a peak-peak distance of 7.5 Oe.

**Figure 4**

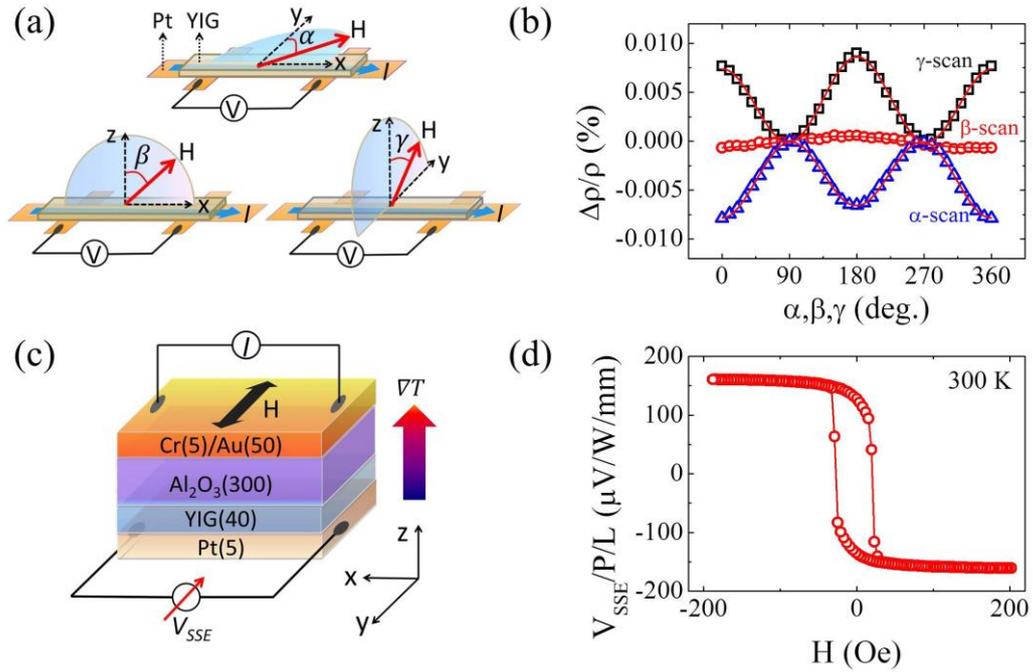

**FIG. 4. SMR and longitudinal SSE of GGG(110)/Pt(5 nm)/YIG(40 nm).** (a) Illustrations of measurement geometry of SMR. α, β and γ are angles between **H** and y, z and z, axes, respectively. The magnitude of **H** is 2000 Oe, 1T, and 1T for α−, β−, and γ−scans, respectively. (b) Angular dependence of SMR ratios for three measurement geometries at 300 K. (c) The sample structure and measurement geometry of longitudinal SSE. The heater current I is 50 mA and **H** is applied along the y direction. All the thicknesses are denoted in nanometers (nm). (d) Field dependence of room temperature SSE signal, which is normalized by the heating power P and detecting length L.